\documentclass[12pt]{article}
\usepackage[pdftex,paperwidth=8.5in,paperheight=11in,top=1in,left=1in,right=1in,bottom=1.2in]{geometry} % letter  8.5
\usepackage{setspace} 
\usepackage{verbatim,amsmath,amssymb,calc,rotating,multicol} 
\usepackage{natbib} 
\usepackage[pagewise]{lineno}
\linenumbers
\usepackage[nomarkers,nolists]{endfloat}

\usepackage[multiple,symbol]{footmisc}

\usepackage{afterpage,caption}

\newcommand{\pd}[2]{\frac{\partial #1}{\partial #2}}
\newcommand{\fiso}{f_\textnormal{iso}}
\newcommand{\opt}{\textnormal{opt}}
\newcommand{\iso}{\textnormal{iso}}

\title{%\todo{check for excessive use of my}
  On the Evolution of Differentiated Multicellularity \vspace{10mm}}

\author{
  Martin Willensdorfer
}

\begin{document}

\maketitle

\begin{spacing}{1.0}
  \centering
  Program for Evolutionary Dynamics\\
  Department of Molecular and  Cellular Biology\\
  Harvard University, Cambridge, MA 02138, USA.  \\
  E-mail: willensd@fas.harvard.edu
\end{spacing}

\vspace{20mm} \noindent
Running Title: Evolution of Multicellularity

\newpage

\begin{center}
  {\large \bf Abstract}
\end{center}

{\bf %
Most conspicuous organisms are multicellular and most multicellular
organisms develop somatic cells to perform specific, non-reproductive
tasks.  The ubiquity of this division of labor suggests that it is
highly advantageous.  In this paper, I present a model to study the
evolution of specialized cells. The model allows for unicellular and
multicellular organisms that may contain somatic (terminally
differentiated) cells.  Cells contribute additively to a quantitative
trait.  The fitness of the organism depends on this quantitative trait
(via a benefit function), the size of the organism, and the number of
somatic cells.  This model allows one to determine when somatic cells
are advantageous and to calculate the optimum number (or fraction) of
reproductive cells. I show that the fraction of reproductive cells is
always surprisingly high.  If somatic cells are very small, they can
outnumber reproductive cells but their biomass is still less than the
biomass of reproductive cells.  Only for non-concave benefit functions can
the biomass of somatic cell exceed the biomass of reproductive cells.
I discuss the biology of primitive multicellular organisms with
respect to the model predictions.  I find good agreement and outline
how this work can be used to guide further quantitative studies of
multicellularity.
}

\newpage
\section{Introduction}

Every organism is exposed to mutations that cause variation in
inherited traits.  Competition between slightly different organisms
leads to the proliferation of variants that increase fitness.  Most
adaptations will fine-tune existing systems but some adaptations lead
to new features.  The evolution of multicellularity was clearly such
an adaptation.  It opened a door to a whole new world of
possibilities~\citep{bonner1965a,book:buss1988a,maynard-smith1995a,book:bonner2001a,book:knoll2003a,nowak2006b}.

In their simplest form multicellular organisms are just clusters of
identical cells.  Such undifferentiated multicellular organisms can
evolve fairly quickly through mutations of surface proteins
\citep{Boraas1998:153,rainey1998a,velicer2003a}.  Cells in such clonal
aggregates do not have to compete against each other for reproduction
since they are genetically identical \citep{book:buss1988a}.  This
alleviation of reproductive competition has a profound effect.  It
allows for a division of labor.  Cells can specialize on
non-reproductive (somatic) tasks and peacefully die since their genes
are passed on by genetically identical reproductive cells which
benefited from the somatic function.  This division of labor turns
multicellular organisms into more than just lumps of cells.  They
contain cells that are different in function and appearance.  Today a
plethora of differentiated organisms exist, demonstrating the
evolutionary success of division of labor.

Most theoretical studies of multicellularity analyze the change in
level of selection and the consequences for reproductive competition
\citep{book:buss1988a,maynard-smith1995a,Michod1997:607,michod1997a,michod2001a}.
In this paper I study which conditions make differentiated
multicellularity desirable.  When does a differentiated multicellular
organism have higher fitness than an undifferentiated or unicellular
organism?  In the model presented here, an organism's fitness depends
on a quantitative trait.  The quantitative trait is determined by the
number and types of cells in the multicellular organism.  The
mathematical model allows one to study which kind of benefits
multicellularity must convey to compensate for its disadvantages.  I
calculate how much (compared to a reproductive cell) a somatic cell
has to contribute to the quantitative trait to make division of labor
advantageous and determine the optimum number/fraction of somatic
cells.

The following section describes the model in detail.  In the Results,
I will first consider the evolution of undifferentiated
multicellularity.  To study the evolution of differentiated
multicellularity I analyze the fitness of organisms of constant size.
Thereafter, I study multicellularity in organisms where the size of
the organism and the fraction of somatic cells is governed by the same
evolutionary forces.  In the Discussion, I use the insights from my
analysis to discuss a broad spectrum of primitive multicellular
organisms.

\section{The model}
In this work I use the rate of biomass production as a measure of
fitness.  The rate of biomass production captures an organism's
ability to grow and reproduce.  It denotes how much new biomass per
unit of existing biomass an organism can generate per unit of time.
For organisms of equal size, production rates are equivalent to
fitness (number of new organisms produced per organism per unit of
time).  The model considers how somatic cells, body size, and benefits
of multicellularity affect the rate of production.  I distinguish
between reproductive and somatic cells but allow for only one kind of
somatic cell.  Somatic cells are different from reproductive cells in
that they are terminally differentiated.  Their biomass does not
contribute to the next generation.  Reproductive cells, on the other
hand, contribute to the next generation.  They can be asexually or
sexually reproductive.

To derive how somatic cells affect fitness, let us first assume that
multicellularity and organism size have no effect on the rate of
production.  In this case, as illustrated in Figure~\ref{fig:model}, a
unicellular organism has the same fitness as a four-cell organism
because four unicellular organisms produce 16 unicellular descendants
(16 cells) after two cell divisions, and one four-cell organism
produces four four-cell descendants (16 cells).  Indeed, if size and
multicellularity have no effect on fitness, then all organisms that
are entirely composed of reproductive cells will have the same
fitness.  Let us use a four-cell organism to derive the cost of somatic
cells.

\begin{figure}[p]
  \includegraphics*[width=\textwidth]{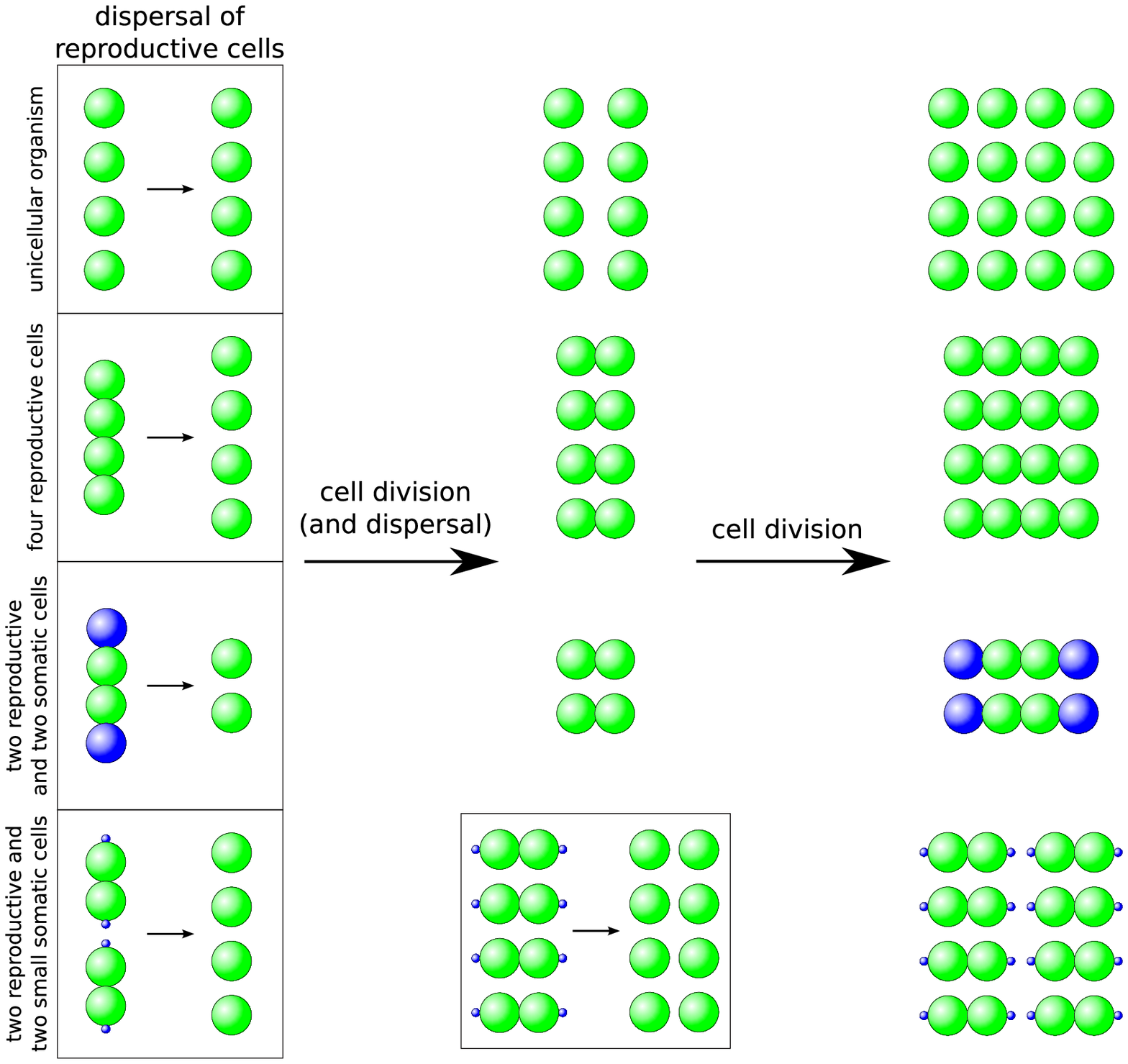}
  \caption{}
\end{figure}
\begin{figure}[p] \ContinuedFloat
  \caption{ \doublespacing%
    Cost of somatic cells in differentiated organisms.  If the rate of
    production is independent of organism size, then every
    undifferentiated organism has the same fitness.  Four unicellular
    organisms produce 16 cells after two cell division (first row), as
    does one four-cell organism (second row).  A four-cell organism
    with two somatic cells (third row) produces half the biomass of an
    undifferentiated organism, because only half its cells are able to
    contribute to the next generation. If the two somatic cells are
    extremely small (fourth row) then nearly all the biomass is
    concentrated in reproductive cells and fitness is equal to the
    fitness of an undifferentiated organism.  In general the fitness
    of a differentiated organism relative to an undifferentiated is
    given by the fraction of biomass that is used for reproduction,
    that is, by $N_r/(N_r + B N_s) = 1- B N_s/(N_r + B N_s)$, where
    $B$ is the size of a somatic cell relative to the size of a
    reproductive cell.}
  \label{fig:model}
\end{figure}

The cost of somatic cells stems from their inability to contribute
directly to the next generation.  To quantify this cost, we can
compare the rate of production of a four-cell organism that has two
somatic cells with the rate of production of a four-cell organism
without somatic cells.  As shown in Figure~\ref{fig:model}, the effect
of somatic cells depends on their size.  If the somatic cells are
negligibly small, then the rate of production of a differentiated
organism is equal to the rate of production of an undifferentiated
organism (forth row in Fig.~\ref{fig:model}).  If somatic cells are as
large as reproductive cells, then a differentiated four-cell organism
with two somatic cells is able to produce only two new four-cell
organisms (8 cells) whereas the undifferentiated organism produces
four new organisms (16 cells).  Hence the rate of production of the
differentiated organism is $1/2$ the rate of production of the
undifferentiated organism. In general, the rate of production
(fitness) is reduced by the fraction of biomass that does not
contribute to the next generation.

In the following we will assign parameters so the biomass and number
of somatic and reproductive cells to quantify this fitness reduction.
I use the Greek letters $\alpha$ and $\beta$ to denote the biomass of
a reproductive and a somatic cell, respectively, in the adult
organism, As we will see, it is sufficient to consider the size of a
somatic cell relative to the size of a reproductive cell.  Let
$B=\beta/\alpha$ denote this size ratio.  Let $N_r$ and $N_s$ denote
the number of reproductive and somatic cells.  An adult organism is
composed of $N=N_r+N_s$ cells and has a body mass of $\alpha N_r +
\beta N_s$.  Of that biomass $\beta N_s$ rests in somatic (sterile)
cells and $\alpha N_r$ in reproductive cells.  Hence, $\beta
N_s/(\alpha N_r+\beta N_s)$ of the organism's biomass is lost in each
generation and constitutes the cost of somatic cells. The fitness of
an organism with somatic cells relative to the fitness of an organism
without somatic cells is given by
\begin{equation}
  1-\beta N_s/(\alpha N_r+\beta N_s) = 
  \alpha N_r/(\alpha N_r+\beta N_s) = N_r/(N_r+B N_s),
\end{equation}
for $B=\beta/\alpha$ as defined above.

So far I have assumed that the rate of production is independent of
the organism size and that multicellularity does not convey benefits.
An overwhelming amount of empirical data shows that the rate of
production decreases with the body mass of an organism
\citep{peters1986a}.  In particular the annual rate of production per
average biomass scales with $W^{-\gamma}$, where $W$ denotes the
average body mass of an adult organism and $\gamma$ is a scaling factor.
This relationship holds from unicellular organisms to mammals, with
body masses ranging from approximately $10^{-10}$ to $10^3$ kg.  As
summarized by \citet[p.\ 134]{peters1986a}, the exponent $\gamma$
might range from $0.23$ to $0.37$.  For small organisms $\gamma$ is
close to $1/4$, the typical allometric exponent of size.  I will
therefore use $\gamma=1/4$ to discuss quantitative results.  From
above, we know that the adult body mass of an organism in this model is
given by $W=\alpha N_r + \beta N_s$.  Hence, the rate of production
decreases by the factor $(\alpha N_r + \beta N_s)^{-\gamma}$.

I model the advantages of multicellularity as a function of the number
of reproductive and somatic cells.  Let $f(N_r,N_s)$ denote this
benefit function.  Therefore, the fitness of a multicellular organism
is given by
\begin{eqnarray}
  \label{eq:F}
  F(N_r,N_s) &=& 
  \textnormal{cost of somatic cells}
  \times \textnormal{cost of size} \times \textnormal{benefit of multicellularity} \nonumber \\
  &=& \frac{\alpha N_r}{\alpha N_r + \beta N_s} \times (\alpha N_r + \beta N_s)^{-\gamma} \times f(N_r,N_s) \nonumber \\
  &=& \frac{\alpha N_r}{(\alpha N_r + \beta N_s)^{1+\gamma}}  f(N_r,N_s) \nonumber \\
  &\propto& \frac{N_r}{(N_r + B N_s)^{1+\gamma}}  f(N_r,N_s),
\end{eqnarray}
where $B=\beta/\alpha$ is the size of a somatic cell relative to the
size of a reproductive cell.

This model confirms common sense.  If multicellularity does not affect
fitness, that is, $f(N_r,N_s) = \textnormal{constant}$, then a
unicellular organism has a higher fitness than a multicellular
organism since $F(N_r,N_s) < F(N_r,0) < F(1,0)$ for $N_r>1$ and
$N_s>0$.  For undifferentiated multicellularity the fitness of an
organism is given by $N_r^{-\gamma} f(N_r,0)$ and multicellularity is
only advantageous if it conveys benefits that compensate for the
disadvantages caused by the size increase.  In other words, $f(N_r,0)$
has to increase more steeply than $N_r^{-\gamma}$ decreases.

Central to my analysis of multicellularity is the function
$f(N_r,N_s)$, which captures the benefit of multicellularity.  I will
assume that somatic and reproductive cells contribute to a
quantitative trait, $x$, and that the benefit of multicellularity is a
function, $f(x)$, of this trait.  For simplicity, I illustrate this
approach by formulating predator evasion and flagellation in terms of
this model.

For predator evasion, the quantitative trait is given by the size of
the organism.  It increases with the number of cells.  Its value
determines to which extent the organism is able to evade predation.
If the organism is big enough, the predator is unable to ingest it and
the benefit of multicellularity, $f(x)$, is close to $1$.  For small
organisms predation might be severe and $f(x)$ close to $0$.  One can
expect a steep increase of $f(x)$ as the organism size surpasses the
maximum particle size the predator can ingest.  Figure~\ref{fig:f}a
shows a benefit function that could be used to describe predator
evasion.

For flagellation, the quantitative trait is given by the flagellar
drive that the cells of the organism provide.  The more cells, the
more flagellar drive, which improves the organism's ability to
maintain its position in a favorable environment.  For this example
the benefit function can be expected to be concave.  An initial
increase in flagellar drive might be very beneficial by allowing the
organism to maintain its position.  At some point, however, the
organism has enough flagellar drive to maintain its position for most
of the time and a further increase in flagellar drive does not yield a
substantial benefit.  Figure~\ref{fig:f}b shows a benefit function
that could be used to model benefits from flagellation.

\begin{figure}[p]
  \centering
  \includegraphics*[width=\textwidth-3cm]{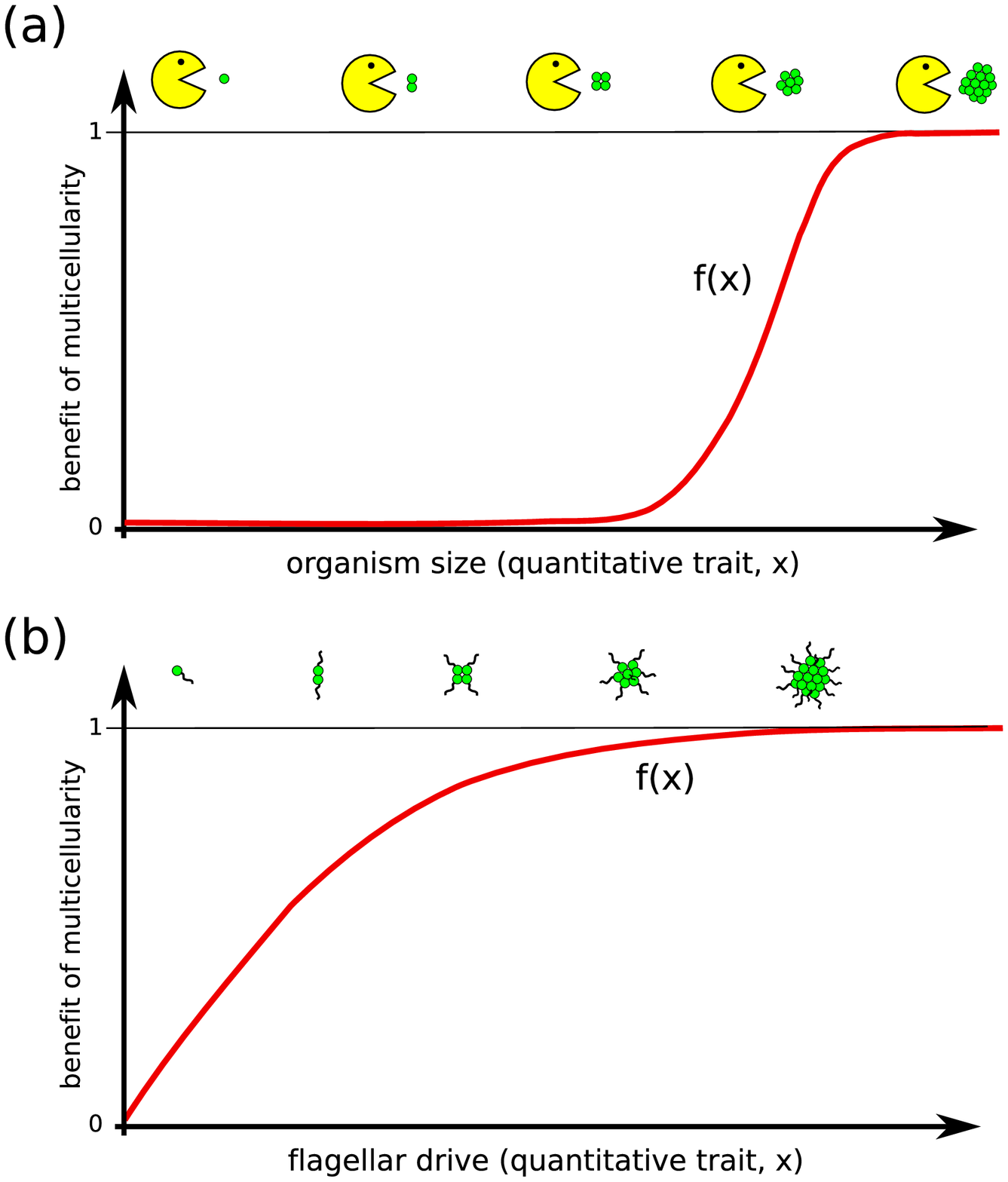}
  \caption{}
\end{figure}
\begin{figure}[p] \ContinuedFloat
  \caption{ \doublespacing%
    The use of benefit functions to model the benefit of
    multicellularity. (a) Predator evasion: Multicellularity can be
    beneficial by allowing the multicellular organism to evade
    predation.  We can model this benefit as a function of the
    organism size (a quantitative trait).  Multicellularity is not
    beneficial if the size of the organism does not exceed the
    predators upper prey size limit. Large benefits can be expected
    for size increases that surpasses this prey size limit.  Once this
    threshold is exceeded and the organism immune against predation,
    further size increases bring little additional benefits.  (b)
    Flagellation: Multicellularity can also increase the motility of
    an organism by increasing its flagellar drive (a quantitative
    trait). Initial increases in motility can bring large benefits
    because they increase an organism's ability to reach favorable
    environments.  At some point, however, a further increase in
    flagellar drive brings little benefits because the organism is
    already able to access the most favorable environment.}
  \label{fig:f}
\end{figure}

As we will see, my analysis of this model does not require an exact
specification of the benefit function.  But it is necessary to make
assumptions about the contribution of individual cells to the
quantitative trait.  

I am mainly interested in the evolution of early multicellular
organisms composed of few cells for which it is reasonable to assume
(at least as an approximation) that cells contribute additively to the
quantitative trait.  I use the letters $a$ and $b$ to denote the
contribution of a reproductive and a somatic cell to the quantitative
trait.  In an organism with $N_r$ reproductive and $N_s$ somatic
cells, the quantitative trait, $x$, is given by
\begin{equation}
  x = a N_r + b N_s.
\end{equation}
Hence, we have
\begin{equation}
  \label{eq:fu}
  f(N_r,N_s) = f(a N_r + b N_s).
\end{equation}
Let $A = a/b$ denote the contribution of a reproductive cell relative
to the contribution of a somatic cell.  Since somatic cells are
specialized cells and free of reproductive constraints, they will
generally contribute more to the trait than reproductive cells
($b>a$), hence, $A = a/b \in [0,1]$.  Furthermore, I consider
organisms where somatic cells are usually smaller than reproductive
cells ($\beta \le \alpha$) and, hence, $B = \beta/\alpha \in (0,1]$.

To simplify the analysis, I scale the argument for $f$ so that $b=1$,
that is, a somatic cell contributes one unit to the quantitative
trait.  By using $A=a/b$ and rescaling the argument for $f$, we can
rewrite (\ref{eq:F}) as
\begin{eqnarray}  
  \label{eq:Fu}
  F
  &\propto& 
  \frac{N_r}{(N_r + B N_s)^{1+\gamma}} 
  f\left (A N_r +  N_s\right).
\end{eqnarray}
Note that if the contribution of reproductive cells to the
quantitative trait is insignificant (i.e., $A N_r + N_s \approx N_s$)
then the benefit of multicellularity is simply a function of the
number of somatic cells.

So far I have not made any assumptions about $f(x)$.  To simplify the
analysis I restrict $f(x)$ to monotone increasing and bound functions.
Both constraints are reasonable. An increase of the quantitative trait
should not lead to a decrease in fitness, and the fitness of an
organism cannot be increased infinitely.  As mentioned earlier, $f(x)$
is scaled such that $\lim\limits_{x \rightarrow \infty} f(x) =1$.

\begin{table}[hp]
  \centering
  \begin{tabular}{lp{12cm}}
    Symbol & Interpretation \\
    \hline
    $N_r$ & Number of reproductive cells \\
    $N_s$ & Number of somatic cells (an organism is 
    undifferentiated if $N_s=0$)\\
    $\alpha$ & Size of a reproductive cell \\
    $\beta$ &  Size of a somatic cell \\
    $B=\beta/\alpha$ & The size of a somatic cell relative to the size of a reproductive cell; $B$ is bound to $[0,1]$\\
    $a$ & Contribution of a reproductive cell to the quantitative trait\\
    $b$ & Contribution of a somatic cell to the quantitative trait\\
    $A=a/b$ & Contribution of a reproductive cell to the quantitative trait relative to the contribution of a somatic cell;  $A$ is bound to $[0,1]$\\
    $x$ & The value of the quantitative trait\\
    $f(x)$ & The benefit function.  It captures the extend to which the organism benefits from the quantitative trait. The benefit function is scaled such that $\lim_{x\rightarrow\infty} f(x)=1$. \\
    $\gamma$ & The allometric exponent which is approximately $1/4$\\
  \end{tabular}
  \caption{Notation summary}
  \label{tab:prm}
\end{table}

Table~\ref{tab:prm} summarizes the variables and parameters used in
this model.

\section{Results}
\label{results}
I am interested in the kind of benefit functions, $f(x)$, that promote
the evolution of multicellularity.  I am also interested in how much a
somatic cell has to contribute to the quantitative trait (relative to
reproductive cells) to compensate for the loss in somatic biomass.  I
calculate the optimum fraction of reproductive cells and the optimum
fraction of reproductive biomass.  This allows us to determine under
which conditions undifferentiated or differentiated multicellularity
evolves and to infer on the composition of early differentiated
multicellular organisms.

First, I analyze conditions for the evolution of undifferentiated
multicellularity.  Thereafter I study the evolution of somatic cells
in organisms of constant size.  I study the unconstrained model last.
In the unconstrained model the size of the organism and the fraction
of somatic cells is governed by the benefit function $f(x)$.  I show
that it is possible to calculate the optimum fraction of reproductive
cells and that this fraction is independent of the benefit function
$f(x)$.

\subsection{Undifferentiated multicellularity}
\label{UndiffMult}
The evolution of undifferentiated multicellularity corresponds to an
evolutionary transition from organisms composed of one (reproductive)
cell to organisms composed of several identical (also reproductive)
cells.  Undifferentiated multicellular organisms have, per definition,
no somatic cells ($N_s=0$).  For $N_s=0$, the fitness (\ref{eq:F})
simplifies to
\begin{equation}
 F(N_r) \propto N_r^{-\gamma} f(N_r,0).
\end{equation}
Multicellularity ($N_r>1$) is only advantageous if $N_r^{-\gamma}
f(N_r,0) > f(1,0)$.  In other words, the benefit function $f(N_r,0)$
has to increase faster than $N_r^{-\gamma}$ decreases.

I employ this simple case to illustrate the method that will be used
to analyze the more complex cases.  I would like to know for which
functions $f(x)$ multicellularity is advantageous and what the optimum
number of reproductive cells is.  To get a general idea of how $f(x)$
affects fitness, we can determine the functions $f(x)$ for which the
fitness is constant with respect to $N_r$.  Let
$f_\textnormal{iso}(x)$ denote these functions.  I refer to them as
isolines since they join points of equal fitness.  They are analogous
to the lines on topographic maps that join points of equal altitude.
These isolines can be used to illustrate the fitness landscape with
respect to benefit functions $f(x)$.  The fitness of an
undifferentiated organism is given by $F=N_r^{-\gamma} f(aN_r)$ and
constant if $f(aN_r) \propto N_r^{\gamma}$.  Substituting $x=aN_r$, we
get
\begin{equation}
  \label{eq:fuisoUndiff}
  \fiso(x) \propto x^{\gamma}.
\end{equation}

\begin{figure}[p]
  \centering
  \includegraphics*[width=\textwidth]{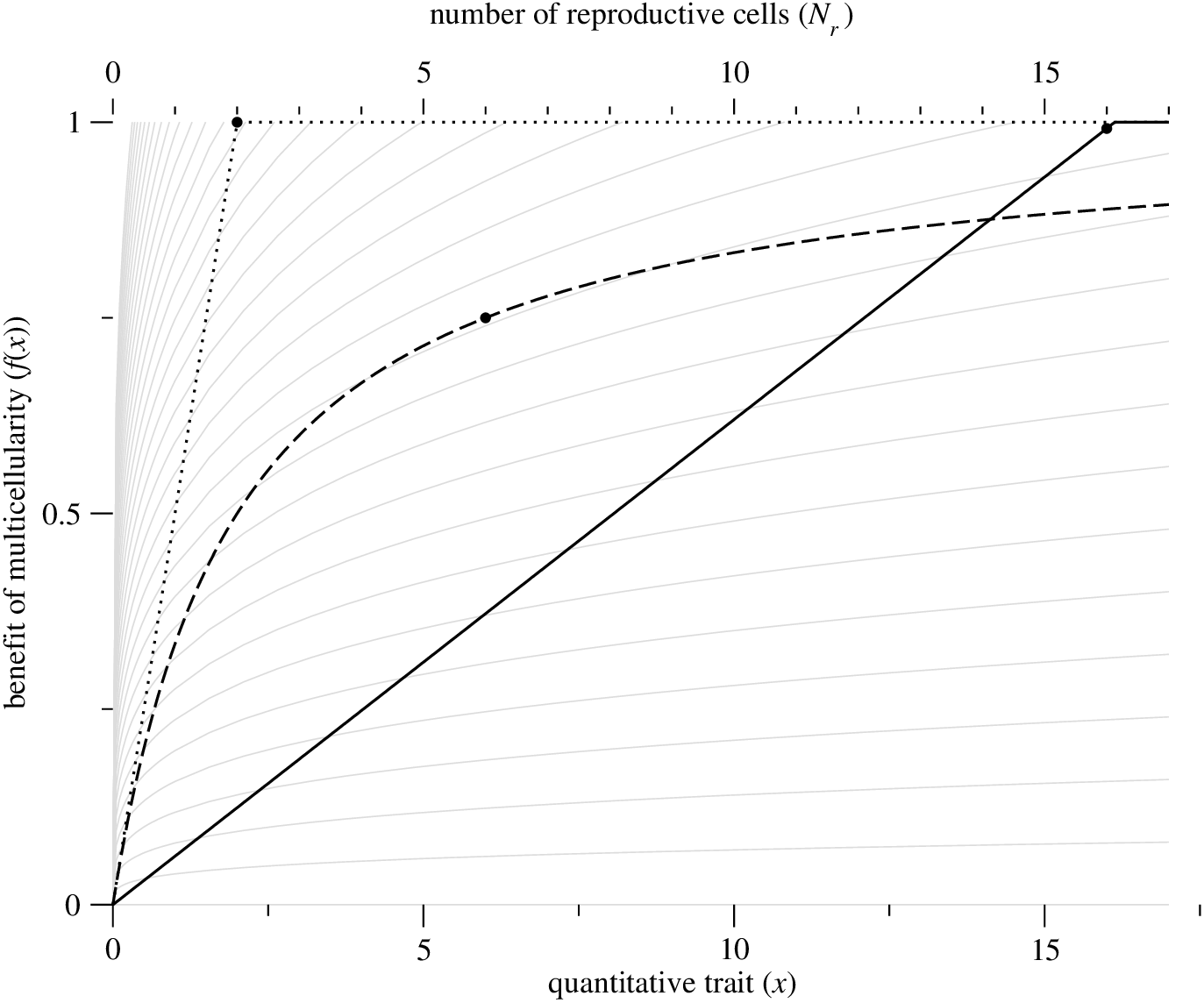}
  \caption{\doublespacing %
    Optimum number of (reproductive) cells in undifferentiated
    organisms.  The gray curves show isolines, i.e., benefit
    functions, $f_\iso(x)=x^{-\gamma}$ (for $\gamma=1/4$), for which
    the fitness (\ref{eq:Fu}) of the organism is constant with respect
    to changes in the number of cells (see Eq.~\ref{eq:fuisoUndiff}).
    The solid and dotted lines show linear benefit functions.  The
    dashed curve shows a concave benefit function.  The number of
    reproductive cells is optimal if $f(x) \le f_\iso(x)$ for the
    isoline with $f(x_\opt)=f_\iso(x_\opt)$, where
    $x_\opt=aN_{r,\opt}$.  Black bullets indicate the optima,
    $(x_\opt,f_\iso(x_\opt))$, for the given benefit functions.}
  \label{fig:fiso_Ns=0}
\end{figure}

The gray curves in Figure~\ref{fig:fiso_Ns=0} show these isolines.
With the knowledge of these isolines it is easy to determine which
functions, $f(x)$, promote the evolution of multicellularity.  It is
also simple to determine the optimum number of reproductive cells.  An
organism with $N_r$ reproductive cells has $x=aN_r$ as value for the
quantitative trait.  We have an optimum $x_\opt = a N_{r,\opt}$ if
$f(x) \le \fiso(x)$ for the isoline with $f(x_\opt)=\fiso(x_\opt)$.
In a continuous setting the optimum satisfies $\pd{}{x}f(x_\opt) =
\pd{}{x}f_\textnormal{iso}(x_\opt)$ for the isoline with
$f(x_\opt)=f_\textnormal{iso}(x_\opt)$.

To interpret isoline plots, it might be useful to keep the analogy
with topographic maps in mind.  One can think of $x$ as the distance
traveled along a particular trail, $f(x)$, in a mountainous region.
The highest point, in our case the optimum, $x_\opt$, is reached if
the trail ``brushes'' the highest contour line along the trail,
$f(x_\opt) = f_\iso(x_\opt)$.  None of the points along the trail will
be above this contour line, $f(x) \le f_\iso(x)$.

Figure~\ref{fig:fiso_Ns=0} shows two linear and one concave benefit
function.  The black bullets indicate the optimum for each benefit
function.  As one can see, a linear benefit function will always
promote the evolution of multicellularity and will increase $f(x)$
until it reaches a value close to one.  A concave benefit function
reaches the optimum earlier and results in smaller organisms.
Undifferentiated multicellularity would not evolve if $f(x)$ increases
slower than the isolines. In particular, multicellularity would not
evolve if $f(a)>2^{-\gamma}$.  For $\gamma=1/4$, we have
$2^{-\gamma}=0.84$ and multicellularity would not evolve if
unicellular organisms are able to benefit from the quantitative trait
more than 84\% of its full potential.

\subsection{Differentiated multicellularity in organisms of constant
  size}
\label{DiffMultConstN}

In the previous section I have shown that undifferentiated
multicellularity is advantageous for many benefit functions.  In the
following I will analyze the evolution of somatic cells
(differentiated multicellularity).  For simplicity, I will first
analyze the evolution of somatic cells in organisms of constant size.

From a biological perspective it is relevant to consider organisms of
constant size, since many benefits of undifferentiated
multicellularity imply constraints on the size of the organism.
Predator evasion, for example, is known to promote the evolution of
undifferentiated multicellularity \citep{Boraas1998:153}.  An organism
that uses multicellularity to evade predation is obviously constrained
with respect to size.  It has to be larger than the largest particle
that the predator can feed on.  Replacing nine large reproductive
cells with nine small somatic cells could decrease its size to
dangerous levels.

For simplicity, I will first assume that somatic and reproductive
cells have the same size ($B=1$) and explore the more general case of
$B<1$ thereafter.

\subsubsection{Somatic cells are as large as reproductive cells ($B=1$)}
The size of an organism is given by $S=\alpha N_r + \beta N_s$.  If
somatic and reproductive cells have the same size
($B=\beta/\alpha=1$), then the size of the organism can only be held
constant if the number of cells that compose this organism is
constant, that is, $N=N_r + N_s = \textnormal{constant}$.  In this
case the fitness, $F(N_r,N_s)$, depends on one variable instead of
two.  Using $N_r$ as this variable we can rewrite (\ref{eq:F}) as
\begin{equation}
  \label{eq:F_NB}
  F \propto 
  N_r f\left [N - (1-A) N_r\right],
\end{equation}
because the trait value of an organism with $N$ cells is given by $x =
N - (1-A) N_r$.  Hence, the quantitative trait of an organism of size
$N$ has a value of at least $AN$ (if $N_r=N$) and at most $N-1+A$ (if
$N_r=1$).  The fitness of an organism is constant if $f\left [N -
  (1-A) N_r\right] \propto N_r^{-1}$.  Expressing the number of
reproductive cells, $N_r$, in terms of the quantitative trait, $x$, we
get
\begin{equation}
  \label{eq:fisoN}
  f_\iso(x) \propto (N-x)^{-1}.
\end{equation}
Since we rescaled $f(x)$ so that a somatic cell contributes one unit
to the quantitative trait, we can interpret $N$ in (\ref{eq:fisoN}) as
the value of the quantitative trait of an organism entirely composed
of somatic cells.  Hence, $N-x$ could be interpreted as the value by
which the quantitative trait is reduced due to the existence of
reproductive cells, i.e., the cost of reproductive cells in terms of
the quantitative trait.

Let us first calculate the optimum number of reproductive cells for a
linear benefit function, $f(x)=cx$.  To guarantee that $0 \le f(x)<1$
for all organisms composed of $N$ cells, we constrain $c$ to
$0<c<\frac{1}{N-(1-A)} \approx 1/N$. As mentioned above, the optimum,
$x_\opt$, satisfies $\pd{}{x}f(x_\opt) = \pd{}{x}f_\iso(x_\opt)$ for
the isoline with $f_\iso(x_\opt)=f(x_\opt)$.  Hence, we have to solve
the two equations $\pd{}{x}f(x_\opt) = \pd{}{x}f_\iso(x_\opt)$ and
$f_\iso(x_\opt)=f(x_\opt)$ for $x_\opt$ and the (irrelevant) constant
$k$ in $f_\iso(x) = k (N-x)^{-1}$.  Solving these equations for the
benefit function $f(x)=cx$, we get $x_\opt=N/2$ which corresponds to
and optimum number of reproductive cells of
\begin{equation}
  \label{eq:fNopt_fu_N=const}
  N_{r,\opt}=\frac{N}{2(1-A)}.
\end{equation}

As we can see, the optimum value for the quantitative trait is
independent of $A$.  The parameter $A$ does, however, determine how
many somatic cells are necessary to reach the optimum value $x_\opt$
for the quantitative trait, and hence $N_{r,\opt}$.  Remarkably, the
optimum number of reproductive cells is independent of the slope, $c$,
of the linear benefit function.  We only constrained the slope so that
the linear function does not exceed one ($f(x) < 1$).  Equation
(\ref{eq:fNopt_fu_N=const}) also shows that the number of reproductive
cells is usually greater than $N/2$ and only equal to $N/2$ if
reproductive cells do not contribute to the quantitative trait
($A=0$).

If the optimum number of reproductive cells, $N_{r,\opt}$, is less
than $N$, then somatic cells are advantageous and differentiated
multicellularity is likely to evolve.  From
(\ref{eq:fNopt_fu_N=const}) we see that this is only the case if
$A<1/2$ which means that somatic cells have to contribute twice as
much as reproductive cells to the quantitative trait to justify their
existence.

\begin{figure}[p]
  \centering
  \includegraphics*[height=\textheight-20mm]{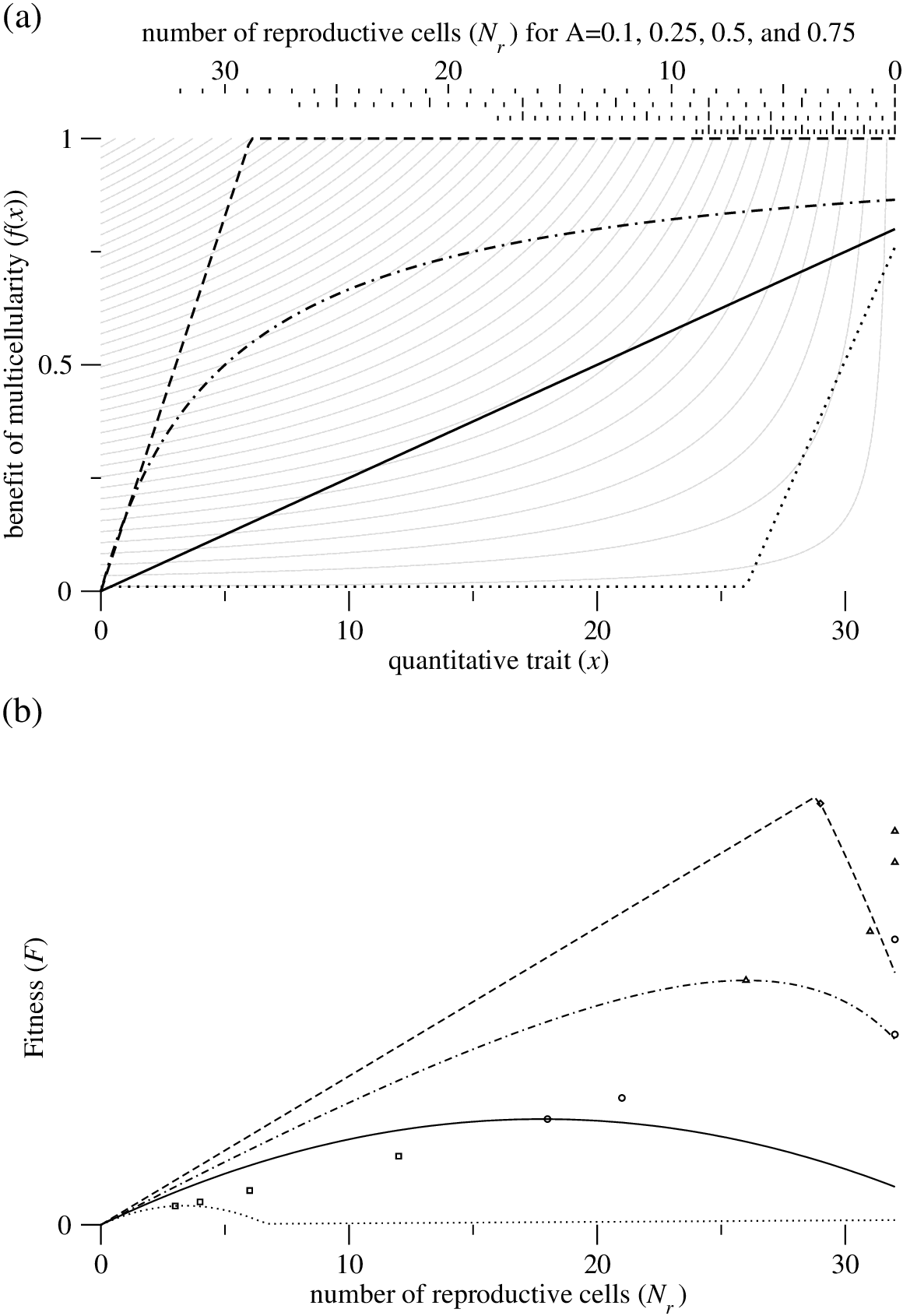}
  \caption{}
\end{figure}
\begin{figure}[p] \ContinuedFloat
  \centering
  \caption{\doublespacing %
    Optimum number of reproductive cells in organisms of constant size
    ($N=32$) with somatic cells that are as large as reproductive
    cells ($B=1$).  (a) The gray curves show isolines with respect to
    changes in $N_r$ (see Eq.~\ref{eq:fisoN}).  Isolines are
    independent of $A$ for $B=1$.  But, since $x=N-(1-A)N_r$, the
    parameter $A$ affects the relation between $N_r$ and $x$.  The top
    axis illustrates how this relation changes with $A$.  (b) Fitness
    of an organism as a function of $N_r$.  The curves correspond to
    the benefit functions from above for $A=0.1$.  To keep the figure
    uncluttered, I do not plot the fitness curves for $A=0.25$, $0.5$,
    and $0.75$, but indicate only the maxima. }
  \label{fig:fuiso_N}
\end{figure}

We can summarize our results for organisms of constant size, uniform
cell sizes, and linear benefit functions: (a) such organisms contain
many reproductive cells, and (b) somatic cells in such organisms have
to contribute substantially more to the quantitative trait than
reproductive cells.  In the following I analyze isoline plots to
illustrate that this result holds for many nonlinear benefit
functions.

Figure~\ref{fig:fuiso_N}a shows four benefit functions and isolines
$f_\iso(x) = (N-x)^{-1}$ for $N=32$.  The solid line represents a
linear benefit function that satisfies the requirements from above
($f(x)<1$).  It is evident that the isolines and the benefit function
have the same slope at $x_\opt = N/2 = 16$.  The Figure illustrates
that $x_\opt$ does not depend on $A$ (the relative contribution of a
reproductive cell to the quantitative trait).  Even though $x_\opt$ is
always with respect to $A$, this parameter determines how many somatic
cells, if any, are required to reach this $x_\opt$.  The top axis
shows how $x$ corresponds to $N_r$ for four different values of $A$.
As $A$ increases to one, the range of possible values for $x$, $AN$ to
$N-(1-A)$, shrinks (to the right).  This is no surprise.  If the
contribution of reproductive and somatic cells to the quantitative
trait are about the same, then the total value of the quantitative
trait changes little if one substitutes a reproductive cell for a
somatic cell.  One can see that if $A$ is larger than $1/2$, then the
quantitative trait of an undifferentiated organism already exceeds the
optimum value of $x_\opt = N/2 < AN$ for a linear benefit function.
There is no need for the organism to evolve somatic cells.

Figure~\ref{fig:fuiso_N}a contains two concave benefit functions
(dashed and dash-dotted curves).  It is easy to localize $x_\opt$ for
these functions and obvious that their $x_\opt$ is smaller than the
$x_\opt$ for the linear benefit function.  For a concave benefit
function, organisms need less somatic cells to optimize fitness and
the functional demand on somatic cells to justify their existence
increases ($A$ has to be even smaller, see upper axis in
Fig.~\ref{fig:fuiso_N}a).  Hence, the results from above, that
organisms have few somatic cells and that somatic cells have to
contribute substantially more to the quantitative trait does also hold
for concave functions.

Examining the isolines in Figure~\ref{fig:fuiso_N}a we see that only a
convex function can lead to organisms with many somatic cells and few
reproductive cells.  A convex function would describe a situation in
which the organism has to obtain a minimum threshold value to benefit
from the quantitative trait.  In such a situation the functional
demand on somatic cells is relaxed and organisms might require many
somatic cells to optimize fitness.

\subsubsection{Somatic cells that are smaller than reproductive cells
  ($B<1$)}

If somatic cells are smaller than reproductive cells, then the total
number of cells, $N$, can change even if the size of the organism
remains constant.  If, for example, reproductive cells are twice as
large as somatic cells ($B=1/2$), then one reproductive cell can be
replaced by two somatic cells, which increases the total number of
cells by one but keeps the organism size constant.

If somatic cells are half the size of reproductive cells, we can
simply apply the results from above by changing the parameter $A$.
Since contributions are additive in my model, somatic cells of size
$\beta$ that contribute $b$ to the quantitative trait are equivalent
to somatic cells of size $\beta/2$ that contribute $b/2$ to the
quantitative trait.  In other words, if somatic cells that are as
large as reproductive cells are beneficial, then somatic cells that
are half as large and contribute half as much to the trait have to be
beneficial as well.  Hence, to get result for $B<1$, we only have to
consider the results from above and replace $A$ with $AB$.  For
example, for a linear benefit function the optimum number of
reproductive cells is given by
\begin{equation}
  N_{r,\opt} = \frac{N}{2(1-AB)}
\end{equation}
and somatic cells are advantageous if $AB<1/2$.  For concave benefit
functions, we can conclude that more than half of the biomass of the
organism will rest in reproductive cells and that somatic cells need
to satisfy $AB<1/2$ to justify their existence. Appendix~A contains a
more technical and detailed analysis of the case $B<1$.  For
completeness, I analyze a model in which the number of cells is held
constant (as opposed to the size of the organism) in Appendix~B.

\subsection{The complete (unconstrained) model}
\label{DiffMult}

Let us now study the unconstrained model.  By not restricting the
number of cells or the size of the organism, I assume that its size
and the optimum fraction of reproductive cells are governed by one
evolutionary force.  In this case the quantitative trait, $x$, can no
longer be expressed as a function of $N_r$.  It depends on $N_r$ and
$N_s$.  This makes the calculation and visualization of isolines
unwieldy.  Instead, we can actually calculate the maximum of the
fitness function.  In the following we will realize that If the size
and the composition of the organism can change freely, then the
optimum fraction of reproductive cells is independent of $f(x)$ and
can be calculated.

Using $N_s=N-N_r$ and $q=N_r/N$ we have
\begin{equation}
  \label{eq:x}
  x=N(1 - (1-A)q)
\end{equation}
and can rewrite (\ref{eq:F}) as
\begin{eqnarray}
  \label{eq:Fflag_noNs}
  F &\propto& \frac{q}{N^{\gamma}[(1-B) q + B]^{1+\gamma}}
  f[N(1 - (1-A)q)].
\end{eqnarray}
In the following I assume that there is at least one reproductive cell
($q>0$) and that somatic cells contribute more to the quantitative
trait than reproductive cells ($A<1$).  Further, I assume that $f(x)$
is differentiable and monotone increasing ($\pd{f}{x}>0$). We can
calculate
\begin{equation}
  \label{eq:pdfN}
  \pd{f}{N}=\pd{f}{x}\pd{x}{N}=\pd{f}{x} \cdot [1 - (1-A)q]
\end{equation}
and
\begin{equation}
  \label{eq:pdfq}
  \pd{f}{q}=\pd{f}{x}\pd{x}{q}=-\pd{f}{x} \cdot N(1-A).
\end{equation}
Since none of the factors in (\ref{eq:pdfN}) and (\ref{eq:pdfq}) equal
zero, we can express $\pd{f}{q}$ in terms of $\pd{f}{N}$,
\begin{equation}
  \label{eq:pdfN2}
  \pd{f}{q}=-\pd{f}{N}\frac{N(1-A)}{1-(1-A)q}.
\end{equation}
Applying the product rule of differentiation to (\ref{eq:Fflag_noNs})
we get
\begin{equation}
  \label{eq:pdFN}
  \pd{F}{N} \propto -\frac{\gamma f}{N} + \pd{f}{N}.
\end{equation}
If the number of cells, $N$, is optimal, then $\pd{F}{N}=0$ and hence
$\pd{f}{N}=\frac{\gamma f}{N}$ which can be substituted into
(\ref{eq:pdfN2}) to give
\begin{equation}
  \label{eq:pdfN3}
  \pd{f}{q}=-\frac{(1-A)\gamma}{1-(1-A)q}f.
\end{equation}
Differentiating (\ref{eq:Fflag_noNs}) with respect to $q$ results in
\begin{equation}
  \label{eq:pdFq0}
  \pd{F}{q} \propto
  [(1-B)q+B] f - (1+\gamma) q (1-B) f + q [(1-B)q+B] \pd{f}{q} 
\end{equation}
Using (\ref{eq:pdfN3}) we can substitute $\pd{f}{q}$ in
(\ref{eq:pdFq0}) and get
\begin{equation}
  \label{eq:pdFq}
  \pd{F}{q} \propto  \left [ B + \gamma q (1-B) - 
    \frac{q ((1-B)q+B)(1-A)\gamma}{1-(1-A)q} \right ] f.  
\end{equation}

The fraction of reproductive cells is optimal if $\pd{F}{q}=0$.  Since
$f>0$, the optimum fraction of reproductive cells can be determined by
calculating for which $q$ the factor in (\ref{eq:pdFq}) equals zero.
Multiplying this equation by $1-(1-A)q$ shows that terms quadratic
in $q$, i.e., $(1-A)(1-B)q^2$ cancel.  Hence, this equation is linear
in $q$ and can be solved to give
\begin{equation}
  \label{eq:qopt}
  q_\opt = \frac{\gamma^{-1}B}{1-\left[1-(1+\gamma^{-1})(1-A)\right ]B}.
\end{equation}

Thus, we are able to calculate the optimum fraction of reproductive
cells, $q_\opt$. Remarkably, $q_\opt$ is independent of the benefit
function $f(x)$.  The benefit function will, however, determine the
size of the organism.

\begin{figure}
  \centering
  \includegraphics*[width=\textwidth]{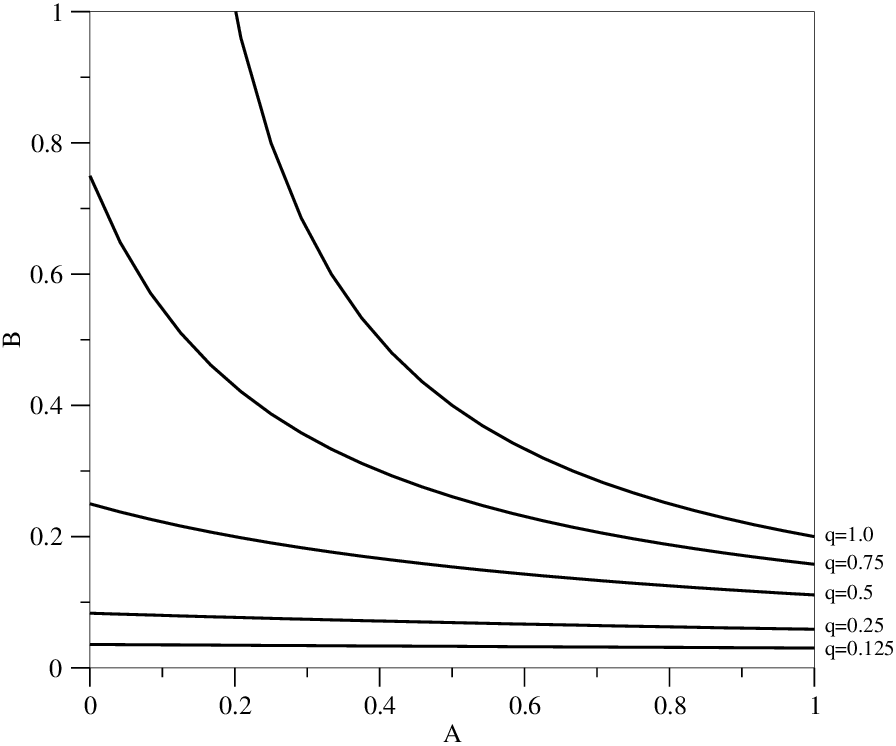}
  \caption{\doublespacing %
    Optimum fraction of reproductive cells.  If the size of an
    organism and the fraction of reproductive cells, $q$, can adjust
    freely, then the optimum fraction of reproductive cells is
    independent of $f(x)$ and given by (\ref{eq:qopt}).  The optimum
    fraction depends on $A$ (the contribution of a reproductive cell
    to the quantitative trait relative to the contribution of a
    somatic cell), $B$ (the size of a somatic cell relative to the
    size of a reproductive cell), and $\gamma$ (the allometric
    exponent).  This implicit plot shows (for $\gamma=1/4$) which
    values $A$ and $B$ result in $q_\opt=1$, $0.75$, $0.5$, $0.25$,
    and $0.125$.  Differentiated multicellularity for an organism of
    size $N$ is beneficial if $q_\opt<1-1/N$.  Note that $A$ becomes
    irrelevant for small $B$.  }
  \label{fig:q_opt_undivisible}
\end{figure}

Figure~\ref{fig:q_opt_undivisible} shows an implicit plot of $q_\opt$
as a function of the parameters $A$ and $B$.  The curve for $q_\opt=1$
gives the threshold for parameter values that favor the evolution of
differentiated multicellularity.  This curve is given by
$AB=\frac{\gamma}{1+\gamma}$ ($=1/5$ for $\gamma=1/4$).  In
particular, if somatic cells are as large as reproductive cells
($B=1$), then they are only beneficial if $A <
\frac{\gamma}{1+\gamma}$.  For $\gamma=1/4$, somatic cells have to
contribute five times as much to the quantitative trait than the
reproductive cell.  Similarly, if somatic cells contribute as much to
the quantitative trait as reproductive cells ($A=1$), then, to be
advantageous, their size has to be a fraction of the size of
reproductive cells. In particular, this fraction has to be less than
$(1+\gamma)/\gamma$.

Figure~\ref{fig:q_opt_undivisible} shows that for small somatic cells
(small $B$), the ability of reproductive cells to contribute to the
quantitative trait ($A$) has little effect on $q_\opt$.  For small
$B$, the denominator in (\ref{eq:qopt}) is approximately 1 and $q_\opt
\approx \gamma^{-1}B$.  From (\ref{eq:Fflag_noNs}) we also see that
$A$ appears in the equation for $F$ only in the term $1-(1-A)q$.  If
$q$ is small (because of small $B$), then $1-(1-A)q \approx 1$ and $A$
has little effect on the fitness of the organism.  

That the effect of $A$ (the ability of reproductive cells to
contribute to the quantitative trait) on $F$ and $q_{r,\opt}$ depends
on $B$ (the size of somatic cells) is an important result for our
understanding of the evolution of somatic cells.  Many reproductive
cells have to grow to a minimum size before they can initiate cell
division.  Newly evolved somatic cells are presumably as large as
reproductive cells but are instantaneously relieved of reproductive
size constraints.  The minimum size at which a somatic cell can still
function might be much smaller than the minimum size of a reproductive
cell.  Organisms will have the tendency to evolve somatic cells that
are as small as possible (decrease $B$).  Equation~\ref{eq:qopt} shows
that a decrease of $B$ increases the optimum number of somatic cells
(decreases $q_{r,\opt}$).  This will further increase the selective
pressure to reduce the size of somatic cells because the organism has
now more somatic cells that should not be unnecessarily large.  This
feedback loop might continue until there are many, small somatic
cells.  At this point (small $B$) the contribution of reproductive
cells to the quantitative trait has no major effect on fitness (see
Fig.~\ref{fig:fuiso_N_B}b).  Hence, reproductive cells can cease to
contribute to the somatic function (the quantitative trait in my
model) with little effect on fitness.  They are free to dedicate their
existence fully to reproductive duties.  Such an evolutionary feedback
loop promotes the evolution of organisms with a strict division of
labor between many, small somatic cells and few, large reproductive
cells.

It is important to emphasize that we treated $q$ and $N$ as continuous
variables.  Especially for small multicellular organisms they are,
however, discrete.  For example, a bicellular organism can have a $q$
of $1/2$ or $1$.  From Figure~\ref{fig:q_opt_undivisible} and
(\ref{eq:qopt}) we know that somatic cells in such a bicellular
organism are (even for $A=0$) only advantageous if they are much
smaller than reproductive cells ($B<\gamma$).  Also, for $B=1$ the
optimum fraction of reproductive cells is always larger than
$1/(1+\gamma)$ ($=4/5$ for $\gamma=1/4$).  Hence, evolutionary
transitions to differentiated multicellularity with somatic cells that
are as large as reproductive cells can only happen in organisms that
are composed of at least six cells.

\newpage
 
\section{Discussion}

I have presented a model for the evolution of undifferentiated and
differentiated multicellular organisms.  In my model three factors
determine an organism's fitness: (a) its size (or biomass), (b) its
investment in somatic (terminally differentiated) cells, and (c) a
quantitative trait that is mainly determined by the number and kind of
cells that the organism is composed of.  The quantitative trait, $x$,
affects the fitness of the organism via a benefit function, $f(x)$
(see Fig.~\ref{fig:model}).  For simplicity I assume that the cells of
a multicellular organism contribute additively to the quantitative
trait.  Since somatic cells are specialized and terminally
differentiated, they can contribute more to the quantitative trait
than reproductive cells.

I analyze under which conditions (benefit function, contribution to
the quantitative trait, size of somatic cells, etc.) the evolution of
undifferentiated and differentiated organisms is favored, and
calculate the optimum fraction of somatic cells. My analysis shows
that undifferentiated multicellularity is favored by many benefit
functions.  The evolution of undifferentiated multicellularity is,
however, unlikely if the unicellular organism is already able to
receive large benefits from the quantitative trait.  In particular,
multicellularity will not evolve if the unicellular organism benefits
from the quantitative trait more than $84\%$ ($=2^{-\gamma}$ for
$\gamma{=1/4}$) of its full potential.

My model suggests that primitive differentiated organisms will
generally have a small fraction of somatic biomass.  If somatic cells
are as large as reproductive cells and the benefit function linear or
concave, then the fraction of somatic cells is always less than or
equal to $1/2$.  If somatic cells are smaller than reproductive cells,
they might occur in large numbers, but their biomass will still be at
most $1/2$ of the total biomass.  Somatic cells compose more than
$1/2$ of the organism only if the benefit function is convex.

In the following I discuss the biology of primitive multicellular
organisms.  First I discuss undifferentiated, then differentiated
organism.  I use experimental data from volvocine algae to demonstrate
how experimental observations can be compared with model predictions
from this work.  At the end I point out the limitations of my model.

Most benefits of undifferentiated multicellularity relate to an
organism's ability to evade predators or its ability to secure a
favorable position in the environment.  Predator evasion is commonly
recognized as a driving force for the evolution of undifferentiated
multicellularity \citep{book:buss1988a,king2004b}.
\citet{Boraas1998:153} showed that unicellular algae can evolve
multicellularity within few generations after exposure to a
phagotrophic predator.  Phagotrophic and many other predators face an
upper size limit for the particles they can ingest.  A simple
``sticking together'' of cells provides protection by exceeding these
size limits.  In this case the quantitative trait is the size of the
organism.  If it exceeds a certain value, the organism benefits
substantially from it (see Fig.~\ref{fig:f}a).

Multicellularity is also known to improve an organism's ability to
obtain a favorable position in the environment.  In particular, the
flagellation constraint dilemma is believed to play an important role
in the evolution of multicellularity \citep{margulis1981a}.  Many
eukaryotic cells face the dilemma that they are unable to maintain
flagellation during cell division and, hence, lose motility
\citep{bonner1965a,margulis1981a,book:buss1988a,Koufopanou1994:907,kirk1997a}.
In an undifferentiated multicellular organism, motility can be
maintained. Multicellularity can also increase the speed of an
organism.  Many cells can provide more drive than a single one
\citep{Sommer1986:650}.  For this example, the quantitative trait,
flagellar drive, determines the organism's ability to reach a
favorable position in the environment which constitutes a benefit (see
Fig.~\ref{fig:f}b).

Multicellularity can also improve an organism's ability to float.
Many algae lack flagella.  They regulate buoyancy through the
production of carbohydrate ballast and/or gas inclusions
\citep{graham1999a}.  Filamentous growth in combination with the
secretion of extracellular polymeric substances allows the formation
of mats that provide a stable structure which can be used to regulate
buoyancy by trapping bubbles \citep{Phillips1958:765,graham1999a}.  In
this case the quantitative trait might be given by the tightness of
the mat.  Tight mats allow to trap many bubbles and allow the cells in
that mat to stay close to the surface water where they receive more
light.

%% Undifferentiated multicellularity can also help to exploit patchy
%% environments.  A network of connected cells, as in fungi, can be used
%% to transport nutrients.  Consider a situation in which one part of the
%% network is a calcium rich area of the environment and another part in
%% a nitrate rich area.  A network of cells can share these resources and
%% benefit from both nutrients.  In this case the resources that
%% multicellularity helps to obtain are apportioned among reproductive
%% cells and are therefore divisible resources according to my definition.
%% My analysis showed that for divisible resources multicellularity needs
%% to be more than the sum of its parts to be beneficial.  This is indeed
%% the case in such networks of cells since cells share resources they
%% would not be able to obtain individually.

Let us now consider differentiated multicellularity.  My analysis
predicts that primitive differentiated multicellular organisms will
generally have many reproductive cells.  More precisely, in most cases
I would expect more than $1/2$ of an organism's biomass to rest in
reproductive cells.  In the following I discuss algae and slime molds,
two groups of organisms for which quantitative data exist.

\begin{figure}[p]
  \centering
  \includegraphics*[width=\textwidth-20mm]{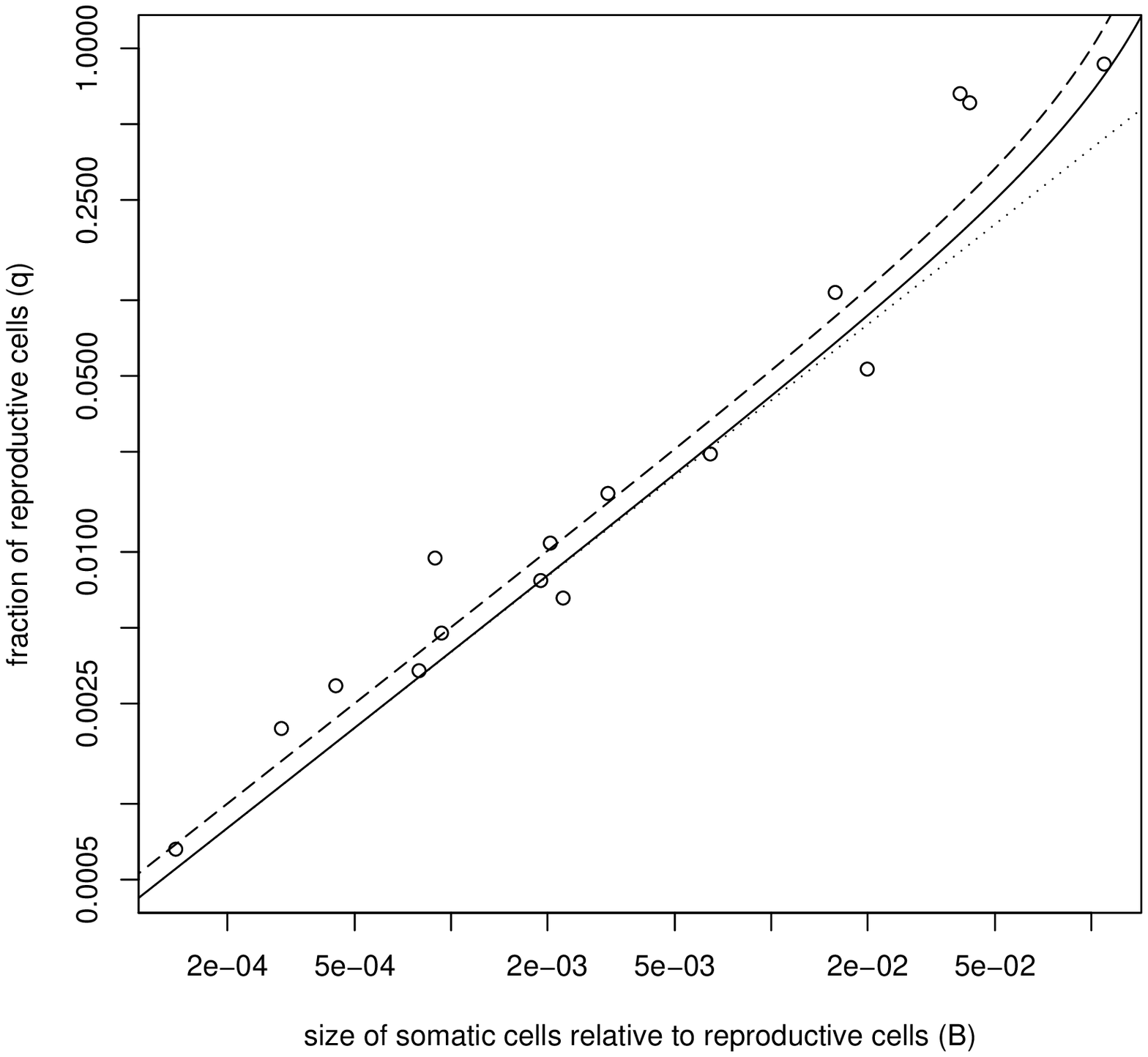}
  \caption{\doublespacing Fraction and size of somatic cells in
    volvocine algae.  I plot the fraction of reproductive cells,
    $q=N_r/N$, as a function of the size of somatic cells relative to
    reproductive cells, $B=\beta/\alpha$.  Points show data collected
    by \citet{Koufopanou1994:907}.  The solid curve shows the model
    prediction (\ref{eq:qopt}) for $\gamma=1/4$ and $A=0$
    (reproductive cells do not contribute to the quantitative trait
    and the benefit of multicellularity is just a largely arbitrary
    function of the number of somatic cells).  An allometric exponent,
    $\gamma$, of 1/5 describes the data best (minimizes the mean
    squared error for data points with $B<3\times 10^{-2}$). The
    dashed line shows the model prediction for $\gamma=1/5$.  The
    dotted line shows $\gamma^{-1}B$, the limit of $q_\opt$ for $B
    \rightarrow 0$.}
  \label{fig:kouf}
\end{figure}

Volvocine algae are an excellent group of organisms to study
differentiated multicellularity.  Their multicellular complexity
ranges from undifferentiated to highly differentiated organisms
\citep{kirk1997a}.  The most primitive differentiated forms have
somatic cells that maintain flagellation during cell division of
reproductive cells.  The flagellar beating is also important to
provide a constant nutrient supply.  It stirs the medium and prevents
a nutrient depletion of the organism's boundary layers which would
occur due to the nutrient uptake by the organism itself
\citep{solari2006a}.  According to the source-and-sink
hypothesis~\citep{bell1985a} somatic cells can also increase the
uptake rate of nutrients, but experiments by \citet{solari2006a}
suggest that the stirring of the medium plays a more important role in
nutrient supply.

The smallest differentiated colonies in volvocine algae have 32 cells
and 4, 8, or 16 are somatic \citep{Goldstein1967:1591,bonner2003a}.
This is in nice agreement with the model prediction.  Allometric data
about soma and germ in volvocine algae shows that there is more germ
tissue than somatic tissue in all species
\citep[Fig.~7]{Koufopanou1994:907}.  Furthermore, we can use the data
collected by \citet{Koufopanou1994:907} to calculate $q$ and $B$.  In
Figure~\ref{fig:kouf}, I compare this data with the model predictions
(\ref{eq:qopt}).  The good agreement between the model and the data
suggests that the size and fraction of somatic cells in volvocine
algae are governed by the same benefit function.

To compare the model predictions with the experimental data, I assumed
that reproductive cells do not contribute to the quantitative trait
($A=0$).  In this case the benefit of multicellularity is modeled as a
(largely arbitrary) function of the number of somatic cells.  To
calculate the optimum fraction of somatic cells, it was not necessary
to specify this function.  My model is in this case surprisingly
general and should be applicable to a wide range of primitive
differentiated multicellular organisms.

Figure~\ref{fig:kouf} suggests that the allometric exponent for
volvocine algae, $1/5$, is smaller than the typical allometric
exponent of $1/4$.  It should be possible to determine experimentally
the allometric exponent of volvocine algae and compare my prediction
with experimental observations.  From experiments similar to the ones
conducted by \citet{solari2006a}, one could also learn something about
the shape of the benefit function by manipulating the number of
(functional) somatic cells.

%% Differentiated multicellularity can also be found in filamentous
%% algae.  Many algae, such as \textit{Cladophora}, evolved rhizoidal
%% cells which are holdfasts that attach the algae to rocks or other
%% substrates close to terrestrial nutrient inputs.  Again, an indivisible
%% resource (favorable position in the environment) and few somatic cells
%% attach several meters of algae filament.

Another organism group that is commonly used to study primitive
differentiated multicellularity is the slime molds.  Slime molds such
as \textit{Dictyostelium discoideum} feed as individual cells until
food becomes scarce, at which point they form a multicellular mass
that migrates to a suitable spot and differentiates into a fruiting
body.  The somatic stalk of the fruiting body lifts the spores above
the ground to facilitate more efficient dispersal
\citep{book:bonner1967a,bonner2003b}.  In this case, the quantitative
trait that conveys the benefit of multicellularity is the height of
the stalk.  The higher the stalk the more efficient is the dispersal.
According to my model, we would expect the biomass of the stalk to be
less then $50\%$ of the total biomass of the fruiting body.
\citet{farnsworth1975a} measured the percentage (dry weight) of stalk
as a function of the temperature during culmination.  This percentage
changes from about 20\% at $18^\circ$C to 13\% at $27^\circ$C.  Hence,
most of the fruiting body is indeed composed of reproductive cells.
The data suggest that somatic cells are slightly more advantageous at
lower temperatures since the fraction of somatic cells increases.
Similarly, in \textit{Myxococcus xanthus}, a fruiting body forming
bacterium, more than $61\%$ of the cells in a fruiting body are spores
\cite[Table 2]{oconnor1991a}.

It would be interesting to collect data about soma and germ in slime
molds that is analogous to the data collected for volvocine algae.  A
comparison of such data with the presented model would be of
particular interest since the quantitative trait is most likely the
size of the stalk.  This trait is easy to measure and would allow
conclusions about the benefit function.  The model presented in this
paper should guide the researcher in their data collection and
presentation.  For example, \citet{Koufopanou1994:907} reported the
average and standard deviation of the number and size of somatic and
reproductive cells.  In the light of my analysis it seems to be of
greater biological importance to report the average and standard
deviation of the fraction of reproductive cells, $q$, and the size of
somatic cells relative to reproductive cells, $B$.

For most conspicuous organism such as plants and animals the number
and biomass of somatic cells vastly outnumbers that of reproductive
cells.  Notably, all of these organisms are much more complex than the
primitive multicellular organisms that are the focus of this study.
They contain many somatic cell types that form organs and interact
with each other in complex ways.  It is important to keep in mind the
kind of organisms that the model is able to describe.  I make two key
assumptions: (a) The benefit of multicellularity can be modeled as a
function of a quantitative trait. In particular, for a given value of
the quantitative trait, the benefit does not depend on the number of
reproductive cells, and (b) that cells contribute additively to that
trait.  If one of these two assumptions is not satisfied, the results
can be quite different.

In Appendix~C I analyze a model in which the fitness of an organism
depends on how much of a limiting resource (e.g., Nitrate) each
reproductive cell obtains.  This resource, after it has been acquired
by the (maybe multicellular) organism, has to be divided between
reproductive cells.  This is in disagreement with assumption (a) since
the benefit depends on the number of reproductive cells (the less
reproductive cells, the more nutrients each reproductive cell
receives).  As shown in Appendix~C, such a situation does not favor
the evolution of undifferentiated multicellularity. Differentiated
organisms will tend to be small and are mostly composed of somatic
cells.

One might also wonder how my model can be applied to organisms for
which the distinction between somatic and reproductive cells is not so
clear cut.  For primitive differentiated multicellular organisms,
``somatic'' cells can be characterized by a delayed cell division or a
reduced probability of reproduction, rather than no cell division or
no reproduction at all.  It is straight forward to incorporate this
developmental plasticity into my model by modifying the term that
captures the cost of somatic cells.  This cost is given by the biomass
that is lost due to the existence of somatic cells (or more generally
the resources that are lost).  For terminally differentiated cells
this is just given by the biomass of the somatic cells but can be
modified to reflect any developmental plasticity.  If, for example,
``somatic'' cells have approximately a $50\%$ chance of reproduction,
then the average evolutionary cost of somatic cells is given by $50\%$
of the somatic biomass.

In this work I mathematically described the costs and benefits of
differentiated and undifferentiated multicellularity.  I showed that
multicellularity can evolve readily if cells of a multicellular
organism contribute additively to a quantitative trait that benefits
the organism in a manner that is independent of the number of
reproductive cells.  Multicellularity is especially beneficial if a
single-cell organism alone cannot benefit from the quantitative trait
substantially.  Only if the single-cell organism is able to exploit
the quantitative trait to $84\%$ ($=2^{-\gamma}$ for $\gamma{=1/4}$)
of the quantitative traits full potential will multicellularity not
evolve.  I showed that evolutionary forces that are based on such
quantitative traits will generally evolve multicellular organism with
few somatic cells even if somatic cells contribute much more to the
quantitative trait than reproductive cells.

In particular, for the complete model (organism size and fraction of
somatic cells is determined by the benefit of multicellularity) and
for somatic cells that are as large as reproductive cells, the optimum
fraction of somatic cells is always less than $\gamma/(1+\gamma)=1/5$.
As a consequence, under such conditions multicellular organisms can
only benefit from somatic cells if they are composed of at least five
cells.  Somatic cells can be numerous if they are very small compared
to reproductive cells but their biomass will still be
less than that of the reproductive cells.  In the presence of many,
small somatic cells, the contribution of reproductive cells to the
quantitative trait has little effect on the fitness of the organism.
This allows reproductive cells to specialize on the reproductive
function and paves the way for a strict division of labor between
reproductive and somatic cells.

\paragraph{Acknowledgments:} I thank Andrew Knoll and David Hewitt for
their biological perspective.  I am grateful to Reinhard B\"urger and
Martin Nowak for comments on the manuscript.  I was supported by a
Merck-Wiley fellowship.  Support from the NSF/NIH joint program in
mathematical biology (NIH grant r01gm078986) is gratefully
acknowledged.  The Program for Evolutionary Dynamics at Harvard
University is sponsored by J. Epstein.

\clearpage
\appendix

\section*{Appendix A: Optimum number of reproductive cells in
  organisms of constant size with small somatic cells.}
In this section I derive the results for organisms of constant size
$S$ and somatic cell that are smaller than reproductive cells ($B<1$).
The unit for $S$ is chosen so that an undifferentiated organism is
composed of $S$ reproductive cells.  The quantitative trait of such an
undifferentiated organism totals $AS$.  A differentiated organism of
constant size with one reproductive cell has $(S-1)/B$ somatic cells
and its quantitative trait equals $(S-1)/B+A$.  Notably, an organism's
quantitative trait can range from $AS$ to $(S-1)/B+A$ and depends on
$A$ and $B$.

For constant size the fitness (\ref{eq:Fu}) is given by
\begin{eqnarray}
  \label{eq:F_u_constS}
  F &\propto& N_r f(AN_r + N_S) \\
  &=& N_r f\left[ S/B - \left(1/B - A\right) N_r \right],
\end{eqnarray}
where we used $N_s = (S-N_r)/B$.  Isolines are given by
\begin{equation}
  \label{eq:fisouS}
  f_\iso(x) \propto (S/B-x)^{-1}
\end{equation}
and are independent of $A$.  The term $S/B$ can be interpreted as
value of the quantitative trait that an organism entirely composed of
somatic cells would have.

For a linear benefit function $f(x) = cx$ we can calculate the optimum
as $x_\opt = S/(2B)$ which corresponds to
\begin{equation}
  \label{eq:NroptS}
  N_{r,\opt}  = \frac{S}{2(1-AB)}.
\end{equation}
Reproductive cells will constitute $1/[2(1-AB)]>1/2$ of the biomass of
the organism.  As for $B=1$, most of the organism's biomass will be
reproductive cells.  Somatic cells are only beneficial if $AB<1/2$.

\begin{figure}[p]
  \centering
  \includegraphics*[height=\textheight-20mm]{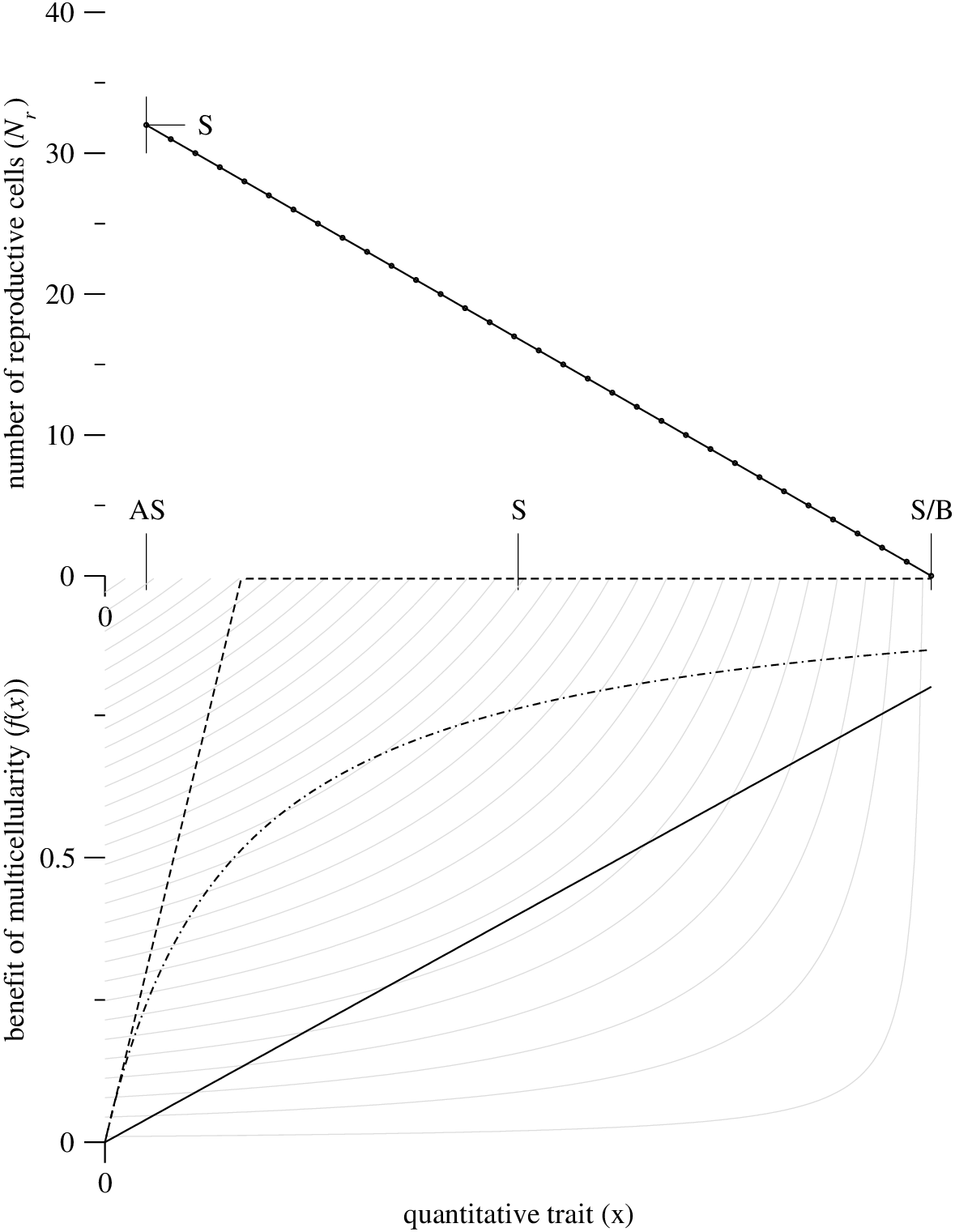}
  \caption{}
\end{figure}
\begin{figure}[p] \ContinuedFloat
  \centering
  \caption{\doublespacing%
    Isolines for organisms of constant size, $S=N_r+BN_s$. The gray
    curves in the lower part of the figure show isolines with respect
    to $N_r$ (see Eq.~\ref{eq:fisouS}).  The top part of this figure
    shows the which $x$ corresponds to which $N_r$. Isolines are
    independent of $A$ and approach $S/B$ asymptotically. Decreasing
    $B$ would increase $S/B$, change the position of the
    isoline-asymptote and the maximum possible value for $x$.  An
    increase in $A$ would increase $AS$ and steepen the line that maps
    $x$ on $N_r$.}
  \label{fig:fuiso_S}
\end{figure}
\afterpage{\clearpage}

Figure~\ref{fig:fuiso_S} shows the isoline landscape and three benefit
functions.  Figure~\ref{fig:fuiso_S}b illustrates how the isolines
depend on the parameter $A$ and $B$.  Figure~\ref{fig:fuiso_S}a shows
how the correspondence between $x$ and $N_r$ depends on $A$ and $B$.
For example, an increase of $A$ would move the upper left end of the
line that maps $x$ on $N_r$ to the right and steepen its slope.
Increasing $B$ would move the lower right point of this line to the
left and also steepen the slope.  It would also change the isolines
which approach $S/B$ asymptotically.  As for constant $N$ most benefit
functions and parameter combinations will lead to a fairly large
number of reproductive cells or a large fraction of reproductive
biomass.  Only for convex benefit functions would $N_{r,\opt}$ be
small.

\section*{Appendix B: Optimum number of reproductive cells in
  organisms with a constant number of cells.}

In this section I analyze the evolution of small somatic cells ($B<1$)
in organisms that are composed of a constant number of cells
($N=$ constant).  The fitness (\ref{eq:Fu}) is given by
\begin{equation}
  \label{eq:F_N}
  F \propto 
  \frac{N_r}{[N_r + B (N-N_r)]^{1+\gamma}} 
  f\left [N - (1-A) N_r\right], 
\end{equation}
and constant if $f\left [N - (1-A) N_r\right] \propto N_r^{-1}[1 + B
(N/N_r-1)]^{1+\gamma}$.  Expressing $N_r$ in terms of $x$, we get
\begin{equation}
  \label{eq:fuisoN}
  f_\iso(x) \propto 
  \frac{[ N - x + B\left( x-AN \right)]^{1+\gamma}}{N-x}.
\end{equation}
We can interpret $N-x$ as the decrease of the quantitative trait due
to the existence of reproductive cells, and $x-AN$ as the increase in
the quantitative trait (compared to undifferentiated organisms) due to
somatic cells.

Let us now study how a change in the size of somatic cells affects the
optimum number of reproductive cells.  The isolines
are given by $1/(N-x)$ for $B=1$. For $B<1$, we can rewrite
equation (\ref{eq:fuisoN}) as
\begin{equation}
  \label{eq:fuisoN2}
  f_\iso(x) \propto 
  \frac{\left[N\frac{1-BA}{1-B} - x \right]^{1+\gamma}}{N-x},
\end{equation}
and notice that $f_\iso(x)$ approaches $(N-x)^\gamma$ for $B
\rightarrow 0$.  The shape of the isoline is entirely determined by
the factor $(1-AB)/(1-B)$ and different combinations of parameters $A$
and $B$ can result in the same isoline.  Since $A$ appears only in the
term $1-AB$, it has less influence on the shape of $f_\iso(x)$ if $B$
is small.  This can be explained intuitively.  If somatic cells are
very small, they are not very costly and how efficient they are
(compared to reproductive cells) is less important.

\begin{figure}[p]
  \centering
  \includegraphics*[height=\textheight-20mm]{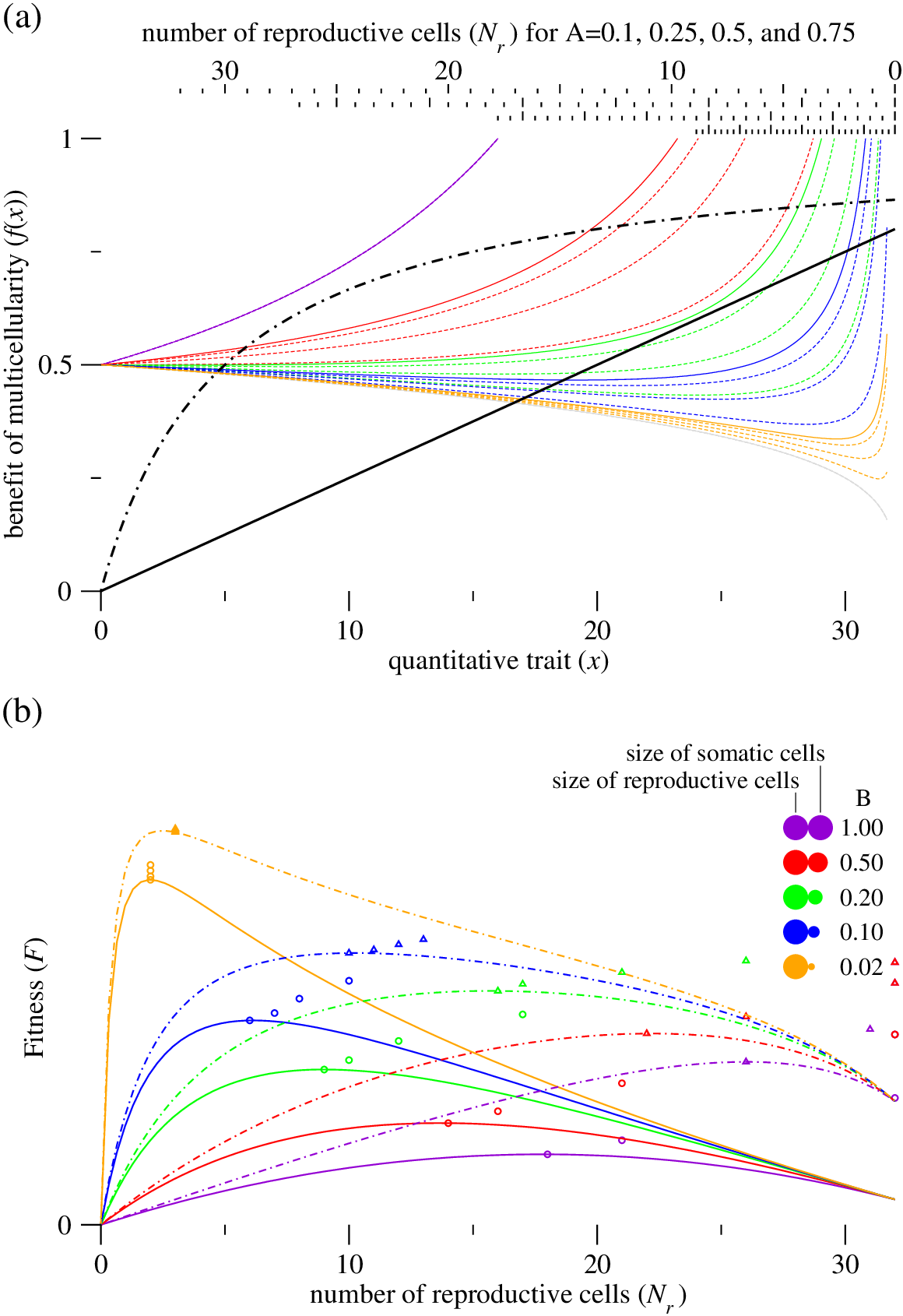}
  \caption{}
\end{figure}
\begin{figure}[p]  \ContinuedFloat
  \centering
  \caption{\doublespacing %
    Optimum number of reproductive cells in organisms with a constant
    number of cells ($N=32$) that have small somatic cells ($B=0.02$,
    $0.1$, $0.2$, $0.5$, and $1$).  (a) Isolines (see
    Eq.~\ref{eq:fuisoN2}) change with $B$ (and $A$) from the purple
    curve ($B=1$) to the gray curve ($B=0$).  I plot isolines for
    $A=0.1$ (solid curves), $0.25$, $0.5$, and $0.75$ (dashed curves).
    The relation between $x$ and $N_r$ is independent of $B$ and shown at
    the top axis.  (b) Fitness of an organism as a function of $N_r$.
    The curves correspond to the benefit functions from above for
    $A=0.1$.  For $A=0.25$, $0.5$, and $0.75$ I mark the maximum for
    each fitness function.  Assuming that cells are spherical, the
    figure legend shows the size differences between somatic and
    reproductive cells for the given $B$ values.  As expected,
    decreasing $B$ decreases $N_{r,\opt}$.  The position and spread of
    the maxima shows that for small somatic cells (small $B$) the
    ability of reproductive cells to contribute to the quantitative
    trait (parameter $A$) has little effect on $N_{r,\opt}$ and the
    fitness of the organism.  }
  \label{fig:fuiso_N_B}
\end{figure}
\afterpage{\clearpage}

Figure~\ref{fig:fuiso_N_B}a shows isolines for different parameter
combinations.  I choose $A=0.1$, $0.25$, $0.5$, and $0.75$, and
$B=0.02$, $0.1$, $0.2$, $0.5$, and $1$.  It illustrates the analytical
results.  Isolines change from $(N-x)^{-1}$ (purple line) to
$(N-x)^\gamma$ (gray line) and different parameter combinations can
result in similar isolines.  The parameter $A$ has little affect on
the isoline if somatic cells are small (small $B$).  The top axes of
Figure~\ref{fig:fuiso_N_B}a show how $x$ corresponds to $N_r$ for
different values of $A$.  Interestingly, if $B$ is small enough, then
isolines can have a negative slope.  In other words, even a constant
benefit function would promote the evolution of somatic cells.  We can
calculate that the slope of $f_\iso(x)$ is negative at $x_{N_r=N}=AN$
if $B<\gamma/(1+\gamma)=1/5$.  For somatic cells of that size the
disadvantage of loosing a reproductive cell is compensated for by the
size decrease (smaller organisms have higher rates of production).  We
will encounter this threshold again during our analysis of the
complete (unconstrained) model.

Figure~\ref{fig:fuiso_N_B}b shows the fitness, $F$, as a function of
$N_r$.  I plot $F$ for $A=0.1$ and indicate the maxima for the other
$A$ values.  As expected, a decrease in $B$ leads to an increase of
$F$ and a decrease of $N_{r,\textnormal{opt}}$.  Also, $F$ and
$N_{r,\textnormal{opt}}$ increase with $A$, but less so if $B$ is
small.

\section*{Appendix C: Multicellularity in organisms in which the
  benefit of multicellularity depends on the number of reproductive
  cells.}

In this section I analyze a small but significant variation of my
model.  To model the benefit of multicellularity, I assumed that this
benefit is a function of a quantitative trait to which somatic and
reproductive cells contribute additively.  In this section I analyze a
model in which the benefit of multicellularity (for a given value of
the quantitative trait) does also depend on the number of reproductive
cells.  In this version reproductive and somatic cells contribute
(additively) to the acquisition of a resource that is desperately
needed by reproductive cells.  The more a reproductive cells has of
this resource, the faster it can grow and the larger is the
probability of survival and, consequently, the fitness of the
organism.  Let $f(x)$ denote the benefit from this resource if the
organism manages to supply \textit{each} reproductive cell with an
amount $x$ of the resource.

If one somatic (reproductive) cell acquires $\beta$ ($\alpha$) of the
resource, then a total of $\alpha N_r + \beta N_s$ can be allocated
between the reproductive cells and each cell would receive
$\frac{\alpha N_r + \beta N_s}{N_r}$.  The benefit of multicellularity
is then given by
\begin{equation}
  \label{eq:f_limited}
  f(N_r,N_s) = f\left( \frac{a N_r + b N_s}{N_r} \right).
\end{equation}
and the fitness of the organism given by
\begin{equation}
  \label{eq:Fl}
  F(N_r,N_s) = \frac{N_r}{(N_r + B N_s)^{1+\gamma}}
  f\left( \frac{a N_r + b N_s}{N_r} \right).
\end{equation}

In the following, I analyze this fitness function.  I show that such
benefits of multicellularity are (a) an unlikely source for the
evolution of undifferentiated multicellularity, (b) differentiated
organisms of constant size would have few reproductive cells, and (c) if
organism size and the fraction of somatic cells are governed by $f(x)$
then the optimum number of reproductive cells is given by one.

\subsection*{Undifferentiated Multicellularity}
It is easy to see that for $N_s=0$, equation (\ref{eq:Fl}) simplifies
to $F = N_r^{-\gamma} f(a)$.  Fitness, $F$, is strictly monotone
decreasing with $N_r$ and optimal for $N_r=1$.  Obviously, if all
cells contributed linearly to the acquisition of a resource and divide
that resource equally among each other, then each cell gets as much as
it would get if it were on its own.  In other words, if cells
contribute linearly to the acquisition of a resource that is
apportioned among them, then multicellularity conveys no advantages
and a unicellular organism has the highest fitness.  Limited resources
that have to be divided between reproductive cells can only trigger
the evolution of multicellularity if cells act synergistically.  The
multicellular organism has to be more than just the sum of its parts.

\subsection*{Constant size}

For constant size $S = \alpha N_r + \beta N_s$, the fitness
(\ref{eq:Fl}) is given by
\begin{eqnarray}
  \label{eq:F_l_constS}
  F &\propto& N_r f(A + N_S/N_r) \\
  & = & N_r f\left[ S/(BN_r) - \left( 1/B -A\right) \right].
\end{eqnarray}
The amount of resources that each reproductive cell receives depends
on $A$ and $B$.  It can range from $x_{N_r=S}=A$ to
$x_{N_r=1}=A+(S-1)/B$.  The isolines are given by
\begin{equation}
  \label{eq:fisolS}
  f_\iso(x) \propto x+1/B-A.
\end{equation}
As one can see, the isolines are linear.  For linear benefit functions
$f(x)=cx$ with $f(S/B)<1$ the optimum number of reproductive cells
equals one if $AB < 1$.  For $B=1$, this simplifies to $A < 1$.  More
so, once differentiated multicellularity is advantageous the optimum
number of reproductive cells is given by one.
Figure~\ref{fig:fliso_S} shows isolines and the non-linear relation
between $x$ and $N_r$ for $A=0.1$.  I plot three possible benefit
functions.  It is obvious that only very steeply increasing benefit
functions would lead to organisms with many reproductive cells.  Most
benefit functions would result in few reproductive cells.

\begin{figure}[p]
  \centering
  \includegraphics*[height=\textheight-20mm]{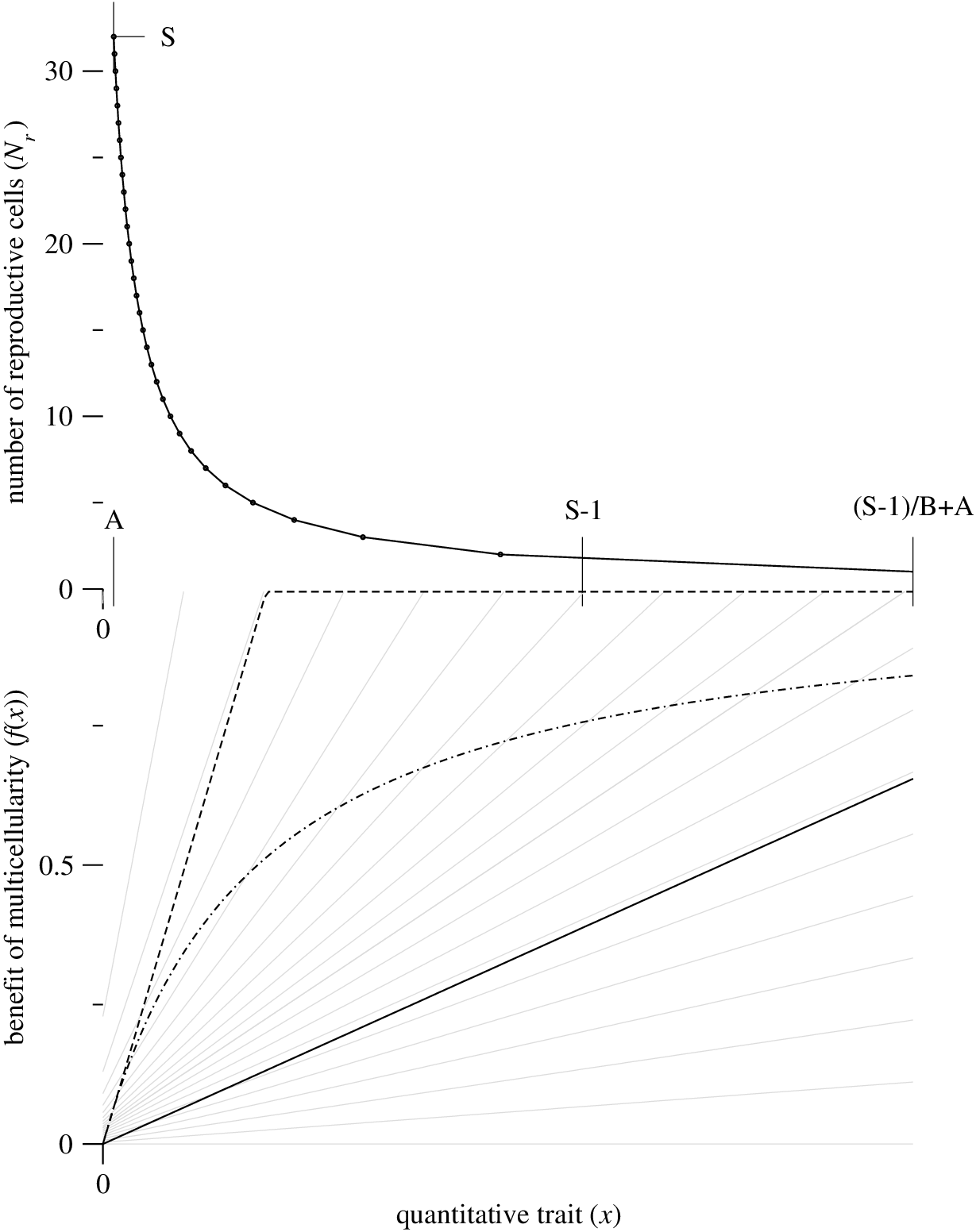}
  \caption{}
\end{figure}
\begin{figure}[p] \ContinuedFloat
  \centering
  \caption{\doublespacing%
    Isolines for organisms of constant size, $S=N_r+BN_s$, that
    benefit from resources that have to be allocated between
    reproductive cells. The gray curves in the lower part of the
    figure show isolines with respect to $N_r$ for $A=0.1$ (see
    Eq.~\ref{eq:fisolS}).  Isolines change with $A$ and $B$ since they
    contain the term $1/B-A$.  The top part of this figure shows how
    $N_r$ changes as a function of $x$. This function depends on the
    parameters $A$ (position of the curve) and $B$ (extension of the
    curve to the right).}
  \label{fig:fliso_S}
\end{figure}

\subsection*{The complete Model}

Let us now analyze the unconstrained model.  As in the main text, the
optimum size of the organism as well as the optimum number of
reproductive cells are governed by the benefit function $f(x)$.  As we
will see, we can use this to our advantage and calculate the optimum
number of reproductive cells analytically.  We can even calculate the
isolines with respect to size and conclude that benefits from
resources that have to be allocated between reproductive cells are an
unlikely cause of the evolution of multicellularity.

The fitness is given by
\begin{equation}
  F \propto N_r( N_r + B N_s)^{-(1+\gamma)} f\left
    (A + \frac{N_s}{N_r}\right ).
\end{equation}
If there is an optimum, it has to satisfy $\pd{F}{N_s}=0$.  We have
\begin{equation}
\label{eqn:pdfFN_s}
\pd{F}{N_s}=N_r\left( -(1+\gamma) B
  (N_r+BN_S)^{-(2+\gamma)}f +  (N_r+BN_S)^{-(1+\gamma)} N_r^{-1} \pd{f}{x}
\right)  
\end{equation}
and hence 
\begin{equation}
  (N_r+BN_S)^{-(1+\gamma)}N_r^{-1} \pd{f}{x} = (1+\gamma) B
(N_r+BN_S)^{-(2+\gamma)}f
\end{equation}
at optimum, which can be used to simplify 
\begin{eqnarray}
\label{eqn:pdfFN_r}
\pd{F}{N_r}&=&(N_r+BN_S)^{-(1+\gamma)}f -N_r(1+\gamma)
(N_r+BN_S)^{-(2+\gamma)} f \nonumber \\
&&- N_s (N_r+BN_S)^{-(1+\gamma)} N_r^{-1} \pd{f}{x} 
\end{eqnarray}
to
\begin{equation}
  \label{eq:pdfFN_rsimple}
  \pd{F}{N_r} = -\left[(1+\gamma)(N_r+BN_S)^2-1\right](N_r+BN_S)^{-(1+\gamma)}f
\end{equation}
which is always negative.  Hence, whenever the fitness is constant
with respect to changes in $N_s$, $N_r$ will decrease.  Since there
can never be less than one reproductive cell the optimum number of
reproductive cells is given by $N_{r,\opt}=1$.  The fitness of the
corresponding organism is given by
\begin{equation}
  \label{eq:Fnutr_N_linf}
  F \propto ( 1 + B (N-1))^{-(1+\gamma)}f[N-(1-A)].
\end{equation}
The isolines with respect to $N$ are given by
\begin{equation}
  \label{eq:fliso_N_full_model}
  f_\iso(x) = \left[ x + 1/B - A \right]^{1+\gamma}.
\end{equation}
Hence, multicellularity based on divisible resources is only
advantageous for convex benefit functions that grow faster than
$\approx x^{1+\gamma}$.  The resulting organism would contain one
reproductive cell and would generally be fairly small (contain few
cells).  This suggests that benefits from resources that have to be
allocated between reproductive cells are an unlikely cause of the
evolution of multicellularity.

\begin{comment}

%%%%%%%%%%%%%%%%
  several modes of asexual production.  Considering the most
  completely examined asexual process, gemmulation, ... survival of
  fluctuating water levels.  Archeocytes accumulate reserve materials
  by phagocytosis of lysed or entire trophocytes.

green and red algae have a continuum of complexity from unicell to
multicell.  Red algae have a more complex life cycle.  Mu

Volvox (green algae)

Lifting (Cladophora p.436), anchoring (Spirogya) , protecting
reproductive cells (Coleochaete p. 533, Fig 21-72 - presumed to be for
nutrient uptake

Bangia atropurpurea (holdfast \& filament)
Ectocarpales (holdfast and filaments

fungi: holocarpic  eucarpic

Nick money (mushroom stem cells)

clonal choanoflagellates (Proterospongia)

\end{comment}

\newpage 

\begin{spacing}{1.0}
\bibliography{papers}
\end{spacing}

\end{document}